# Pushing the Study of Point Defects in Thin Film Ferrites to Low Temperatures Using In-situ Ellipsometry

*Yunqing Tang, Francesco Chiabrera*, Alex Morata, Iñigo Garbayo, Nerea Alayo, Albert Tarancón**

Y. Tang, Dr. F. Chiabrera, Dr. A. Morata, Dr. Iñigo Garbayo, Dr. N. Alayo, Prof. A. Tarancón
Department of Advanced Materials for Energy Applications, Catalonia Institute for Energy Research (IREC), Jardins de les Dones de Negre 1, 08930, Sant Adrià del Besòs, Barcelona, Spain.

Dr. Iñigo Garbayo, Centre for Cooperative Research on Alternative Energies (CIC energiGUNE), Basque Research and Technology Alliance (BRTA), Alava Technology Park, Albert Einstein 48, 01510 Vitoria-Gasteiz, Spain

Prof. A. Tarancón
ICREA, Passeig Lluís Companys 23, 08010, Barcelona, Spain.

E-mail: f.chiabrera@gmail.com ; atarancon@irec.cat

Unveiling point defects concentration in transition metal oxide thin films is essential to understand and eventually control their functional properties, employed in an increasing number of applications and devices. Despite this unquestionable interest, there is a lack of available experimental techniques able to estimate the defect chemistry and equilibrium constants in such oxides at intermediate-to-low temperatures. In this study, the defect chemistry of a relevant material such as $La_{1-x}Sr_xFeO_{3-\delta}$ (x = 0.2, 0.4 and 0.5) was obtained by using a novel in-situ spectroscopic ellipsometry approach applied to thin films. Through this technique, the concentration of holes in $La_{1-x}Sr_xFeO_{3-\delta}$ (LSF) was correlated to measured optical properties and its evolution with temperature and oxygen partial pressure was determined. In this way, a systematic description of defect chemistry in LSF thin films in the temperature range from 350 to 500ºC was obtained for the first time, which represents a step forward in the understanding of $La_{1-x}Sr_xFeO_{3-\delta}$ (x = 0.2, 0.4 and 0.5) for emerging low temperature applications.

1. Introduction

Transition metal oxides thin films have been largely investigated in different strategic research fields such as electronics, solid state ionics and ionotronics due to their wide variety of fascinating functional properties.[1,2] In this family of materials, point defects, imperfections and higher dimensional extended defects are known to severely impact the overall functional properties.[3,4] As a matter of example, oxygen vacancies were shown to enhance oxygen conductivity in oxide-ion electrolytes[5] or to boost oxygen incorporation and catalytic activity in mixed ionic electronic conductors (MIECs)[6,7] while weakening electronic and magnetic order in ferromagnetic oxides.[8] Besides, the presence of heterogeneous and homogenous interfaces in thin films was shown to drastically impact defect concentrations in such layers, which gives rise to deviations from bulk defect chemistry and, eventually, to new and unexpected properties.[9–11] Therefore, the knowledge and quantification of the chemical reactions that dominate the defect concentration in oxide thin films (*i.e.* defect chemistry) is essential for understanding the material behaviour and for engineering their properties. This is especially relevant at intermediate-to-low temperatures (below 500 ºC), where a high electrochemical activity and the nanometric dimensions of the thin films allows the point defect equilibrium with the environment,[12] hampering the use of the high temperature defect chemistry model from the bulk counterpart.

A paradigmatic example of the large effect of point defects on the functional properties of transitional metal oxides can be found in the $La_{1-x}Sr_xFeO_{3-\delta}$ (LSF) model family. LSF compounds crystalize in a perovskite $ABO_3$ structure and find application in many renewable energy technologies, such as Solid oxide Fuel Cells (SOFC),[13] or electrochemical and photo-electrochemical water splitting.[14,15] In LSF, the substitution of trivalent La by divalent Sr gives rise to the generation of electronic holes and/or oxygen vacancies for electronic compensation, depending on the electrochemical equilibrium with the oxygen partial pressure of the environment. Interestingly, both point defects were found to be strongly correlated to many functional properties of LSF. For instance, the increase of holes concentration was found responsible for a large modification of the electronic structure,[16] affecting not just the electronic and magnetic transport properties[17] but also the oxygen evolution properties in aqueous media.[15] Meanwhile, oxygen vacancies were shown to take part into the rate-limiting step of oxygen incorporation at high temperature.[18] For these reasons, the development of a reliable and flexible in-situ method for tracking the point defects of LSF thin films on any substrate and environment is fundamental for tailoring their properties.

Although many different techniques are available for the measurement of the defect concentration in bulk oxides (*e.g.* thermogravimetry, coulometric titration...), the reduced mass and thickness of thin films pose severe challenges in their applicability to nanometric-thick layers.[19] The majority of *in-situ* methods available for deducing oxygen equilibrium mechanisms in oxides thin films are based on an indirect probe of point defects, *i.e.* a physical property dependent on one or more defects is measured as a function of the oxygen activity and the defect equilibrium is obtained by fitting a defect chemistry model. For instance, the measurement of the electrical conductivity as a function of oxygen partial pressure (*i.e.* Brower analysis) is expected to follow the behavior of the majority charge carrier (electrons or holes), which can then be used to fit a proper defect chemistry model to obtain the equilibrium constants. [20,21] The major drawback of the Brower conductivity method is that the charge carrier mobility may also vary with the oxygen partial pressure, giving rise to a non-trivial interdependence difficult to untangle. Another common method for getting insights into the oxygen nonstoichiometry of thin films is measuring the lattice parameters by X-ray diffraction (XRD), since a unit-cell expansion is commonly observed in oxygen deficient oxides.[22]

Nevertheless, challenges in the application of bulk atomic lattice refinement models and the presence of substrate-induced misfit strain hinder the exact quantification of the point defect concentration, making XRD a suitable method mainly for a qualitative analysis of the oxygen deficiency. The measure of the chemical capacitance as a function of oxygen partial pressure is also a very interesting method to extract the defect chemistry of oxide thin films.[23,24] In this technique, electrochemical impedance spectroscopy (EIS) is performed to extract the chemical capacitive contribution of the electrodes, which can be related to the defect species involved in the oxygen incorporation by the application of a proper defect chemistry model. The main advantage of the method is the possibility of modelling the defect chemistry in situ and under real electrochemical conditions (*i.e.* under electrochemical bias), which was used, for instance, to get mechanistic information about the relationship between bulk point defects and oxygen incorporation and evolution reactions.[18] Nevertheless, the measurement of chemical capacitance by EIS is problematic at low temperature, where the high polarization resistance (in parallel to the chemical capacitance) shifts the impedance to very low frequencies, increasing the error associated to the measurement.

A very promising family of methods for tracking point defects in oxides thin films are UV-visible optical characterization techniques.[4,25] As recently reviewed by Buckner et al.[26], point defects strongly modify the electronic band structure of many oxides, giving rise to unique changes of optical properties that can be tracked by optical methods. Although UV-visible techniques cannot directly probe the concentration of point defects, by measuring the evolution of the dielectric properties of the thin film by in-situ Brower analysis it is possible to indirectly achieve a quantification of the defect chemistry of the material, similar to the measurement of the electronic conductivity but without the limitation of a non-constant charge carrier mobility. In this regard, optical transmission was used to quantify the defect concentration and the kinetics of oxidation/reduction of many different oxides thin films, such as Pr doped ceria[25] and La and Fe doped $SrTiO_3$.[27] However, the use of a transmission mode entails a non-trivial drawback, since the substrate must be transparent to light, what limits the application of this technique to specific cases or to limited photon energy windows. Finally, it was recently shown that Raman spectroscopy can also be used for tracking the oxygen non-stoichiometry of $SrTi_{1-x}Fe_xO_{3-\delta}$ thin films.[28]

Taking into consideration the benefits and limitations of the previous methods, an in-situ ellipsometry technique is proposed here for quantifying the defect chemistry of $La_{1-x}Sr_xFeO_{3-\delta}$ (LSF) (x=0.2, 0.4 and 0.5) thin films through the measurements of optical conductivity. Ellipsometry is an optical spectroscopic technique based on the measure of the variation of polarization of a light beam reflected on a thin film sample. It is a non-destructive technique that can be used to measure many structural and optical characteristics of a thin film, such as optical constants and thickness, independently on the optical properties of the substrate. In this work, ellipsometry spectra are acquired under different electrochemical conditions and as a function of temperature (T = 350 - 525 °C) for $La_{0.8}Sr_{0.2}FeO_{3-\delta}$ (LSF20), $La_{0.6}Sr_{0.4}FeO_{3-\delta}$ (LSF40) and $La_{0.5}Sr_{0.5}FeO_{3-\delta}$ (LSF50) thin films grown on Yttria-stabilized Zirconia (YSZ) substrates. The low energy transition of LSF optical conductivity is straightforwardly related to the concentration of electronic holes, which is then measured for different equivalent oxygen partial pressures generated by ion-pumping in an electrochemical cell. In this way, the defect chemistry of LSF thin films is unveiled in an unprecedented low temperature range, demonstrating the great capabilities of the in-situ ellipsometry approach. Moreover, the results show the importance of non-dilute interaction among point defects in LSF thin films, largely influencing the energetics of the oxygen equilibrium reaction.

## 2. Results and discussion

### 2.1. Microstructural and morphological analyses

LSF thin films with different Sr concentration (LSF20, LSF40 and LSF50) were prepared by Pulsed Laser Deposition (PLD) on top of gadolinium-doped ceria (CGO)-coated YSZ (001) substrates. The CGO layer (~10 nm) was deposited on YSZ substrates as a barrier to prevent the formation of secondary phases at the LSF/YSZ interface.[29,30] **Figure 1a** shows the X-ray diffraction (XRD) patterns of the as-deposited multilayers. The LSF20 thin film shows a pseudo-cubic polycrystalline structure with two main orientations along the (h00) and (hk0) directions, while LSF40 and LSF50 films present a single (h00) preferential orientation. The increase of the Sr content in the LSF layers gives rise to a shrinkage of the out-of-plane lattice parameter, as shown by the progressive shift towards higher angles of the (100) diffraction peaks in **Figure 1b.** This contraction of the unit cell with the Sr content is consistent with literature data (see **Figure S1** in the Supplementary Information), and is associated to a subsequent partial oxidation of the Fe ions, being the ionic radius of the $Fe^{4+}$ smaller than the one of the $Fe^{3+}$.[31-33] The structural differences among the samples are also confirmed by the surface topography measured by Atomic Force Microscopy (AFM) (see **Figure 1c-e**). Here, a nanocrystalline structure is clearly observed for the LSF20 (rms = 0.9 nm), while a smoother surface was measured for the LSF40 and LSF50 layers (rms = 0.3 nm). The increase of structural order in the samples with higher Sr content is due to the decrease of the lattice parameter observed rising the dopant concentration, progressively approaching the half diagonal of the in-plane lattice constant of the CGO (3.85 Å) and allowing a textured growth.[34,35] Due to the quite large thickness of the thin films deposited (around 100 nm, see Experimental section) the textured LSF40 and LSF50 layers are not expected to be subjected to misfit elastic strain imposed by the substrate, as also confirmed by the good match of the lattice parameter with bulk references (see **Figure S1** in the Supplementary Information). It must be noted here that the presence of high angle grain boundaries in the LSF20 thin film may locally modify the point defect concentration, as was previously found in other MIEC materials (see Section S.3. in Supplementary Information for the discussion about the effects of homogeneity of the thin films and its effects on the ellipsometric parameters).[9][36] Overall, it is possible to conclude that the deposited LSF layers provide a representative set of samples for the study of the defect chemistry in LSF thin films.

**2.2 Hole effects on the optical conductivity of LSF**

Electronic holes strongly modify the optical absorption and the bandgap of the LSF family.[16,37,38] Although this effect is well documented in literature, to the best knowledge of the authors, a direct relationship between hole concentration and optical features is still lacking. For this reason, as a first step, here we study the optical properties of LSF thin films at room temperature by spectroscopic ellipsometry. In order to remove all the oxygen vacancies and observe the true effect of electronic holes on the optical absorption, the samples were fully oxidized by electrochemical method at 400 ºC (see experimental section for more details). Under these conditions, the concentration of holes in the films is entirely determined by the Sr content and the electronic equilibrium can be written as:

$$[Sr_{La}'] = [Fe_{Fe}^\bullet] = x \quad (1)$$

Where $[Sr_{La}']$ and $[Fe_{Fe}^\bullet]$ are the concentration of Sr and localized electronic holes represented according to the Kröger-Vink notation. Moreover, LaFeO$_3$ (LFO) thin films with no formal concentration of holes (x = 0) were also measured for comparison. The resulting optical conductivity spectra measured at room temperature are shown in **Figure 2a** (please, refer to Supplementary Information **Section S.2** to see the ellipsometry raw data and the description of the five-Lorentzian oscillators model employed in the fitting process [32]). Observing the optical conductivity, one can clearly note that increasing the hole doping leads to strong modifications of the optical properties of the material, consisting in an increase of spectral weight of the low energy transitions around 1 eV and 3 eV (named A and B, respectively) and a decrease of the high energy features around 4.5 eV (labelled C). These results are in good agreement with previously reported works, which deeply analysed the origin of optical transitions in the LSF system by means of different spectroscopy techniques.[15,39–41] According to these studies, the parent LaFeO$_3$ presents a semiconductor structure in which the valence band is mainly composed by hybridized O 2p-Fe e$_g$ orbitals and the conduction band by empty minority spin Fe t$_{2g}$ states, while changing the Fe oxidation state in LSF through Sr doping introduces new intra-bandgap states, as depicted in **Figure 2c**. The exchange of electrons between all these bands have been directly correlated to the A, B and C features observed in the optical conductivity spectra (**Figure 2c**). More specifically, transition A corresponds to the electron promotion from the valence band to the intra-bandgap states induced by hole doping while transition B and C are assigned to the transfer of electrons from O 2p-Fe e$_g$ orbitals to Fe t$_{2g}$ states and from deeper electronic states towards the empty minority spin Fe e$_g$ orbitals[42], respectively. In the energy range considered, the Sr doping mainly affects the optical transitions indirectly by modifying the concentration of holes in the system (appearance of new Fe intra-bandgap electronic states, **Figure 2c**) and not directly, since available Sr electronic states are far from the Fermi level.[16,37,38]

For the purpose of investigating LSF´s defect chemistry, transition A was analysed in more detail since this feature is correlated with the concentration of holes in the doped system.[43,44] For this analysis, the contribution of the transition A to the total spectra was deconvoluted (see **section S.2** in Supplementary Information) showing a linear relationship between the associated maximum of the optical conductivity and the concentration of holes in the system (**Figure 2b and inset**). This linear relationship holds also at high temperature, although a different slope is observed due to the effect of temperature on the electronic band structure[42] (see **Figure S4** in the Supplementary Information). Moreover, the fully reduced LSF50 samples ($[Fe^{\bullet}_{Fe}]\sim 0$) shows an optical spectra similar to the LFO layer and no low energy transition A (see **Figure 1a**), confirming that electronic holes are at the origin of the changes observed in the optical properties. It must be noted here that the linear relationship found here may not hold if LSF films present misfit strain, structural defects or other phenomena able to modify both the electronic band structure and the energetics of oxygen incorporation in layers. Moreover, grain boundaries or dislocations could introduce local modification of the electron holes concentration, introducing uncertainties in the quantification of the optical conductivity (see Section S.3. in Supplementary Information for the discussion about the effects of homogeneity of the thin films and its effects on the ellipsometry parameters). Nevertheless, the strong experimental and theoretical relationship between transition A and the electron holes in LSF suggests that ellipsometry may be used to track the defect chemistry of LSF thin films.

**2.3 In-situ ellipsometry measurements as a function of temperature and equivalent $pO_2$**

We then investigated the variation of optical conductivity in LSF thin films under real electrochemical conditions. In particular, it is possible to measure the material under test as an electrode of an electrochemical cell. If this cell is employed as an oxygen pump against the material, a wide range of oxygen partial pressures can be covered by simply applying a voltage bias.[28,45,46]

In this work, Ag/YSZ/LSF electrochemical cells were fabricated and afterwards measured in a special setup with electrical probes and a heating stage designed for in-situ/operando spectroscopic ellipsometry analysis (**Figure 3a inset** and experimental section). As depicted in the inset of **Figure 3a**, DC voltage bias ($\Delta V$) was applied between the LSF layer and the silver counter electrode both in anodic ($\Delta V>0$) and cathodic ($\Delta V<0$) modes. In cathodic mode, the voltage forces the gaseous oxygen to be reduced and incorporated into the LSF while, in anodic mode, the cell operates in reverse way, excorporating oxygen at the thin films' surface. This electrochemical bias also modifies the oxygen chemical potential of the LSF layer, varying the equivalent oxygen partial pressure experienced by the material ($p_{O_2}^{LSF}$) according to the Nernst potential (the reader is strongly encouraged to see **Section S.5.** in the Supplementary Information, where a detailed discussion of the estimation of the $p_{O_2}^{LSF}$ and the fulfilment of required homogeneous oxygen chemical potential in the LSF layers is presented).

In-situ spectroscopic ellipsometry measurements of LSF thin films were then carried out under different electrochemical bias, i.e. equivalent oxygen partial pressures, as a function of temperature (T = 350 - 525 ºC). **Figure 3a** shows the optical conductivity of the LSF50 samples at 400 ºC for different equivalent $pO_2$ (refer to **section S.5.** in supplementary information for the ellipsometry raw data). One can note that, decreasing the oxygen partial pressure, a gradual decrease of spectral weight of transition A and B takes place, along with an increase of transition C, giving rise to a radical change of optical properties of the LSF thin film, which also changes colour from black to almost transparent. The variation of optical conductivity lowering the $pO_2$ is similar to the behaviour observed in the fully oxidized samples when decreasing the Sr content (see **Figure 2a**), indicating that the electronic holes progressively deplete in the LSF thin film. This hypothesis is also supported by the evolution of the low energy transition A (**Figure 3b**), gradually reducing the spectral weight of the hole-induced empty states above the Fermi level.[38] It must be noted here that all the measurements were perfectly reversible and reproducible (see **Figure S6** in supplementary information). This means that the changes observed in the in-situ measurements are entirely originated by the variation of hole concentration and not by structural features (*i.e.* micro voids, cracks) that would lead to a strong non reversible behavior.

Motivated by the linear relation found in the previous section between transition A and the hole concentration in the LSF system, the maximum of the optical conductivity of A was plotted as a function of the oxygen partial pressure (Brouwer diagram), see **Figure 3c**. The behaviour found is in agreement with the diluted defect model developed by Mizusaki *et al.*[47], in which the holes progressively decrease their concentration following the oxygen incorporation reaction:

$$V_O^{\bullet\bullet} + \frac{1}{2}O_2 + 2Fe_{Fe}^{\times} \leftrightarrow 2Fe_{Fe}^{\bullet} + O_O^{\times} \tag{2}$$

With diluted equilibrium constant ($K_{ox}^{di}$):

$$K_{ox}^{di} = \frac{[Fe_{Fe}^{\bullet}]^2 \cdot [O_O^{\times}]}{(pO_2)^{\frac{1}{2}} \cdot [V_O^{\bullet\bullet}] \cdot [Fe_{Fe}^{\times}]^2} \tag{3}$$

Here $[Fe_{Fe}^{\times}]$, $[V_O^{\bullet\bullet}]$, $[Fe_{Fe}^{\bullet}]$ and $[O_O^{\times}]$ refer respectively to the concentration of $Fe^{3+}$, oxygen vacancies, $Fe^{4+}$ holes and oxygen ions written according to the Kröger-Vink notation. Note that considering both the range of oxygen pressure and the temperature conditions in this work, the concentration of $Fe^{2+}$ electrons are expected to be negligible,[23,47] so that the electronic equilibrium in the LSF system can be simplified to:

$$[Sr'] = 2[V_O^{\bullet\bullet}] + [Fe_{Fe}^{\bullet}] \tag{4}$$

**Equation (3)** and **(4)** can be used to fit the evolution of hole concentration with the $pO_2$ found by spectroscopic ellipsometry, see **Figure 3c.** The diluted model well describes the experimental data for the LSF50 sample, endorsing the possibility of optically measuring in-situ the concentration of holes in the system under real operation conditions. It is interesting to note here that ellipsometry is also very sensitive to the thickness of the layers and may therefore be used for tracking the expansion of the LSF layers taking place during the samples reduction. Although the trend observed for LSF50 layer is in agreement with literature measurements of the chemical expansion coefficient[31] (see **Figure S10** in Supplementary information), the concomitant large variation of the optical properties observed in our case reduces the sensitivity of the thickness measurements, hindering the confidence of the fitting. Nevertheless, it is important to remark that this approach might be used to study the defect concentration of oxides in systems where the optical properties (electronic structure) are not modified upon oxygen reduction, simply by tracking their chemical expansion/contraction.[48]

The point defect concentration as a function of equivalent $pO_2$ was then measured with the same procedure for LSF thin films with different Sr content. **Figure 4** shows the evolution of the electronic holes in the layers obtained by fitting the maximum of optical conductivity at 400 °C. One can note that, decreasing the Sr concentration, the dilute model starts to fail in describing the hole concentration, especially at intermediate $pO_2$, where the films are constantly more reduced than expected. In other words, increasing the hole concentration the oxygen incorporation becomes progressively more difficult, giving rise to a less steep growth of $[Fe_{Fe}^{\bullet}]$. Similar non-ideal behaviours were reported in literature for other oxides, such as $La_{1-x}Sr_xCrO_{3-\delta}$,[49] $La_{1-x}Sr_xCoO_{3-\delta}$,[50,51] $SrFeO_{3-\delta}$,[52] $Ba_{1-x}La_xFeO_{3-\delta}$,[53] and are commonly described considering the Gibbs free energy of incorporation reactions (**Equation 2**) defined by a standard term ($\Delta G_{ox}^{id}$ constant for any $pO_2$) and an activity term ($\Delta G_{ox}^{ex}$), as:

$$\Delta G_{ox} = -RTln(K_{ox}) = \Delta G_{ox}^{id} + \Delta G_{ox}^{ex} \qquad (5)$$

The activity term $\Delta G_{ox}^{ex}$ represents the deviation from the standard free energy of the ideal solution, which is the driving energy for the modification of the oxygen incorporation equilibrium (see **Section S.7.** in supplementary information). A commonly employed model for describing non-dilute behaviour in oxides was proposed by Mizusaki et al., who considered $\Delta G_{ox}^{ex}$ to be linearly proportional to the point defect concentration, as:[49]

$$\Delta G_{ox}^{ex} = a[Fe^{\cdot}] \qquad (6)$$

In this work we found that a quadratic approximation better fits the experimental data obtained ($\Delta G_{ox}^{ex} = b[Fe^{\cdot}]^2$), probably due to a non-negligible interaction between the point defect concentration (see **Figure S11** in Supplementary information). The fitting based on this model leads to a satisfactory description of the experimental data, see **Figure 4**. The ideal equilibrium constant $K_{ox}^{id}$ and quadratic non-ideal parameter $b$ obtained at 400 ºC for the different thin films is shown in **Figure 5a**. In agreement with the literature, the oxidation constant progressively lowers increasing the Sr content, shifting the chemical equilibrium of **Equation (2)** towards a more reduced state. However, also the quadratic parameter $b$ is observed to present roughly the same behaviour, rapidly decreasing with the Sr content, similarly to what was observed in $La_{1-x}Sr_xCrO_{3-\delta}$.[49]

A comprehensive model for describing the physical origin of $\Delta G_{ox}^{ex}$ and its relation with the other point defects is still uncertain. Lankhorst *et al.* used a rigid band model formalism to interpret the variation of $\Delta G_{ox}$ measured in $La_{1-x}Sr_xCoO_{3-\delta}$.[50,51] In their model, the non-ideal behaviour was explained by a rigid shift of Fermi level when increasing/decreasing the electron concentration, giving rise to a modification of the enthalpy of oxygen incorporation. For metallic and semi-metallic oxides, an alternative explanation based on hole degeneracy was also proposed, where the oxygen incorporation leads to the formation of delocalized and highly degenerated holes that modifies their activity.[54,55] Although a modification of the electron hole activity may be at the origin of the behaviour observed in the $La_{1-x}Sr_xFeO_{3-\delta}$ thin films, the presence of a finite bandgap, especially for low Sr content, and of localized electronic defects makes these two models not applicable.[43] Nevertheless, the large variation of the optical conductivity observed reducing the hole concentration, along with the strong modification of band structure and energy levels measured in previous works,[16,37] suggests a strong non-ideal behaviour of the hole chemical potentials, which is expected to affect both the enthalpy and the entropy of $\Delta G_{ox}$. Moreover, the trend observed for the non-ideal parameter as a function of Sr content (see **Figure 5a**), suggests that the non-dilute behaviour is related to the larger bandgap of these films, while increasing the dopant concentration the bandgap tends to reduce and more semi-metallic behaviour is observed, decreasing the influence of the band structure changes on $\Delta G_{ox}$. The most interesting direct consequence of this behaviour is that the concentration of oxygen vacancies is consistently higher than the one predicted by the ideal behaviour, especially for the LSF20 thin film. Although the derivation of the exact mechanism behind the non-ideal behaviour observed is out of the scope of this work, the results shows the unique capabilities of spectroscopy ellipsometry to sense the hole concentration in LSF thin films and to disclose non-dilute phenomena in the defect chemistry of MIEC oxides at intermediate-to-low temperature. Finally, the evolution of the equilibrium constant $K_{ox}^{id}$ with the temperature was obtained (see **Figure S12** in supplementary information), and is plotted in an Arrhenius-like representation in **Figure 5b**. Here, equilibrium constants derived from previous high temperature measurements are also reported and extrapolated to the low temperature range investigated in this work.[46,56] A good agreement is observed between this set of data and our measurements, despite the difference of temperature window and of the nature of samples, suggesting that difference commonly found in thin films may not be due to the energetics of the oxygen incorporation equilibrium but to deviation from ideal dilute systems, more predominant at low temperature.

## 3. Conclusions

The defect chemistry of LSF thin films with different Sr content was studied by spectroscopic ellipsometry concluding that low energy transitions (~1eV) observed in the optical conductivity can be used for tracking the concentration of holes in the material. Moreover, using LSF as an electrode in an electrochemical cell coupled to an ellipsometer, this approach can be extended to calculate the concentration of holes as a function of temperature and oxygen partial pressure in a wide range from $10^{10}$ to $10^{-17}$ bar. In this way, the defect chemistry of LSF20, LSF40 and LSF50 thin films was unveiled for the intermediate-to-low temperature range, not accessible with other techniques. The equilibrium constants extracted from the analysis of the optical constants variation shows a good agreement with the extrapolation from high temperature bulk measurements. Nevertheless, deviations from the dilute defect model are observed in the LSF layers, especially for low Sr contents, consisting in a less steep increase of hole concentration while oxidizing the sample. Overall, in this work, it was possible to push the current limits for the direct measurement of defect chemistry in $La_{1-x}Sr_xFeO_{3-\delta}$ (x=0.2, 0.4 and 0.5) thin films to lower values of temperature, which is crucial for a systematic description of transition metal oxides nowadays in the core of several emerging energy and information technologies.

## 4. Experimental section

*Thin film deposition:* $La_{1-x}Sr_xFeO_{3-\delta}$ (x= 0.2, x=0.4 and x=0.5) thin films were deposited by pulsed laser deposition (PLD) on 10 nm CGO/YSZ (001) substrates. Commercial pellets of LSF20 and CGO were used as target materials, while the pellets of LSF40 and LSF50 were prepared by solid state synthesis. For the home-made targets, first $La_2O_3$, $SrCO_3$ and $Fe_2O_3$ powders were stoichiometrically mixed in an agate mortar and heated up to 1250 ºC for 12 h in air. Then, the synthesized LSF powder was uniaxially pressed into a one-inch pellet and sintered in air at 1300 ºC for 12 h. All the layers were deposited employing a large-area system from PVD products (PLD-5000) equipped with a KrF-248 nm excimer laser from Lambda Physik (COMPex PRO 205). The LSF films were grown with an energy fluency of 0.8 J cm$^{-2}$ per pulse at a frequency of 10 Hz. The substrate was kept at 700 °C, in an oxygen partial pressure of 0.0067 mbar during the deposition and the substrate–target distance was set to 90 mm. The CGO barrier layer was deposited under the same conditions but at a temperature of 750 °C. All the thin films were deposited using a microfabricated Si mask that allowed the deposition of 2 mm x 3 mm rectangular layers at the centre of the 10 mm x 10 mm YSZ (001) substrates.

*Thin film Characterization:* Microstructural characterization and phase identification were carried out by XRD in a coupled Ѳ-2Ѳ Bragg-Brentano configuration using a Bruker D8 Advanced diffractometer equipped with a Cu Kα radiation source. Cross-section image of the as-deposited film was characterized using a Scanning Electron Microscopy (SEM) (Zeiss Auriga) and elemental composition of the LSF thin films was analysed using a SEM-coupled EDX Spectroscope (Zeiss Auriga). The analysis shows a Sr content of 0.17±0.01, 0.34±0.01 and 0.44±0.01 for the LSF20, LSF40 and LSF50 thin films, respectively. Topography of the LSF thin films was characterized in non-contact mode by AFM of XE 100 model provided by Park System Corp.

*Electrochemical characterization:* Silver paste was used on the backside of the YSZ substrate as counter electrode. For increasing the current collection, gold paste was painted on the 2 mm x 3 mm sides of the LSF thin films. On one corner of the YSZ electrolyte, a silver reference electrode was painted, which was used to check the potential applied across the YSZ electrolyte (see inset of **Figure 3a**). The measurements were carried out at intermediate temperatures (490 ºC, 440 ºC, 395 ºC and 350 ºC) using a heating stage (Linkam instruments THMS350), which not only heats up the samples but also includes electrical contacts for in-situ electrochemical measurements during the experiment. The LSF thin films were electrochemical analysed by a staircase potentio-electrochemical impedance spectroscopy (SPEIS) using a potentiostat from Biologic (model SP-150). A DC voltage bias from 0.3 V to -0.6 V was applied at a step of 0.05 V between the gold current collector and the back-side Ag counter electrode. Once the current was stabilized, the electrochemical impedance spectra were recorded with an AC voltage of amplitude 0.01 V in a range of frequency from 1 MHz to 0.1 Hz. All the experiments were carried out in atmospheric air.

*Spectroscopic ellipsometry measurements:* The optical constants of the LSF thin films were measured by spectroscopic ellipsometry employing an ellipsometer (UVISEL, Horiba scientific) in a photon energy range from 0.6 eV to 5.0 eV with an interval of 0.05 eV. The angle of incident light beam was 70 º. The ellipsometry data were modelled and fitted using DeltaPsi2 software from Horiba scientific (see supplementary information for the description of the fitting procedure). The ellipsometry measurements show the thickness of 94±1 nm, 98±1 nm and 112±1 nm for the LSF50, LSF40 and LSF20 thin films, respectively. In-situ ellipsometry measurements were carried out during the electrochemical characterization after each voltage bias step. The ex situ optical conductivity of the fully oxidized LSF samples was obtained by heating the samples at 400 ºC and applying a 0.3 V anodic bias across the YSZ electrolyte. The fully reduced ex situ optical conductivity of the LSF50 layer was instead obtained by applying -0.4 V cathodic bias at 400 ºC. The temperature of the samples was then rapidly decreased continuously applying the anodic bias (assuring a fully oxidized state) down to room temperature, where the ellipsometry spectra were recorded.

**Supporting Information**

Supporting Information is available from the Wiley Online Library or from the author.

**Acknowledgements**

This research was supported by the funding from the European Research Council (ERC) under the European Union's Horizon 2020 research and innovation programme (ULTRASOFC, Grant Agreement number: 681146). This research has also received funding from the NANOEN project (2017 SGR 1421). This project has received funding from the European Union's Horizon 2020 research and innovation program under grant agreement No 824072 (HARVESTORE).

Received: ((will be filled in by the editorial staff))

Revised: ((will be filled in by the editorial staff))

Published online: ((will be filled in by the editorial staff))

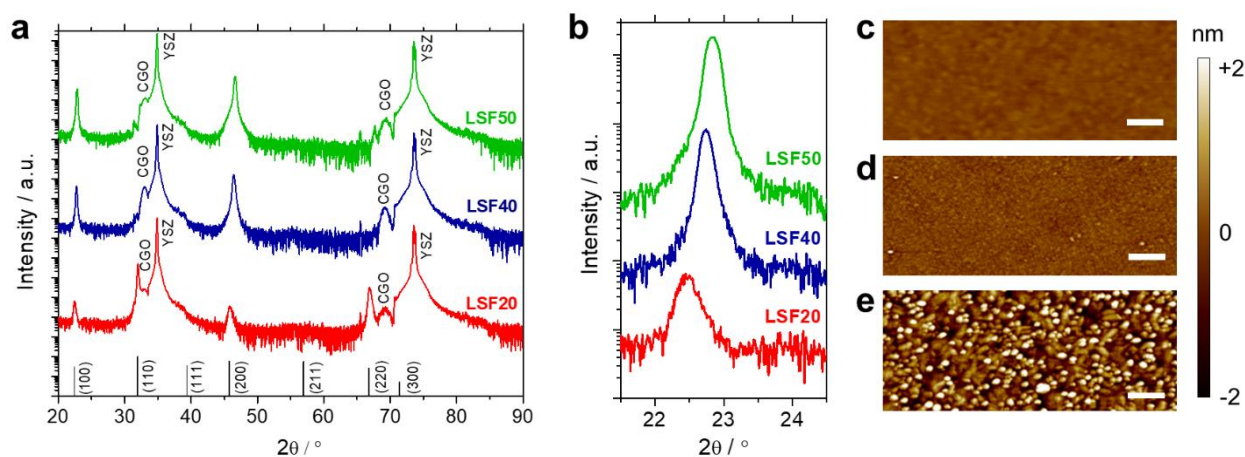

**Figure 1.** a) XRD patterns of the as-deposited LSF20, LSF40 and LSF50 thin films on the (100)-oriented YSZ substrate with the CGO interlayer. The CGO and YSZ phases are labelled while the LSF phase is identified by indicating the reflection peaks; b) Detail of the (100) diffraction peak showing the shift towards higher 2θ angles while increasing the Sr content; AFM images of the as-deposited c) LSF50 d) LSF40 and c) LSF20 thin films. The scale bar in the images is equal to 250 nm.

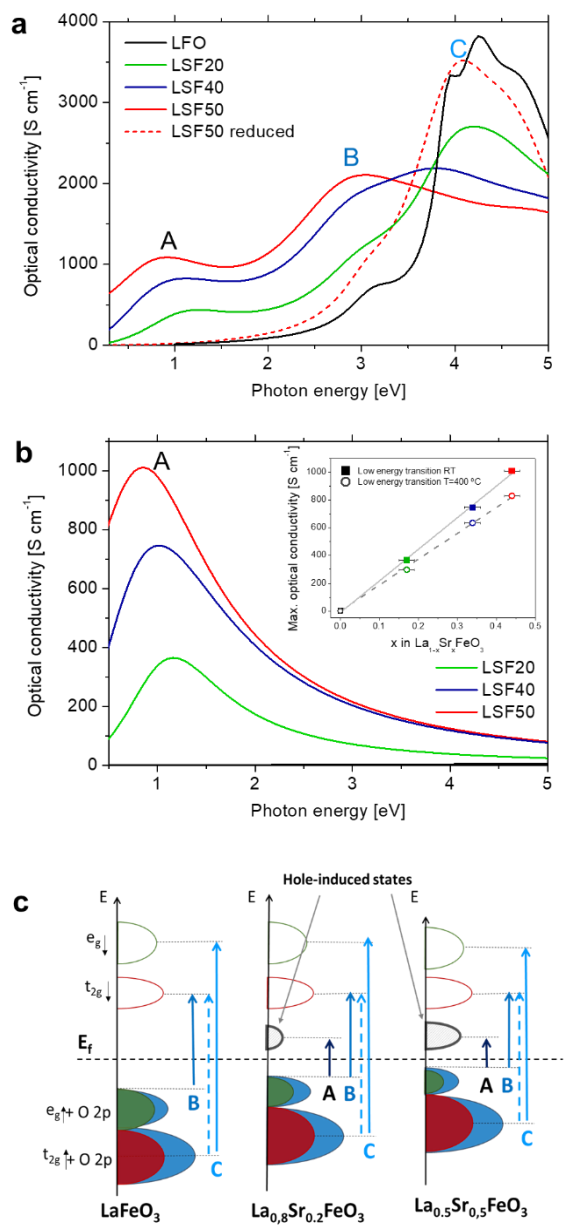

**Figure 2.** a) Optical conductivity spectra of the fully oxidized LFO, LSF20, LSF40 and LSF50 thin films measured by spectroscopic ellipsometry at room temperature. b) The deconvoluted optical conductivity of the low energy transition A. The inset shows the linear relation between the maximum optical conductivity and the hole concentration in the LSF system at room temperature and 400 ºC. c) Sketch of the LSF electronic band structure and main optical transitions.[32,33,37,42,43,57].

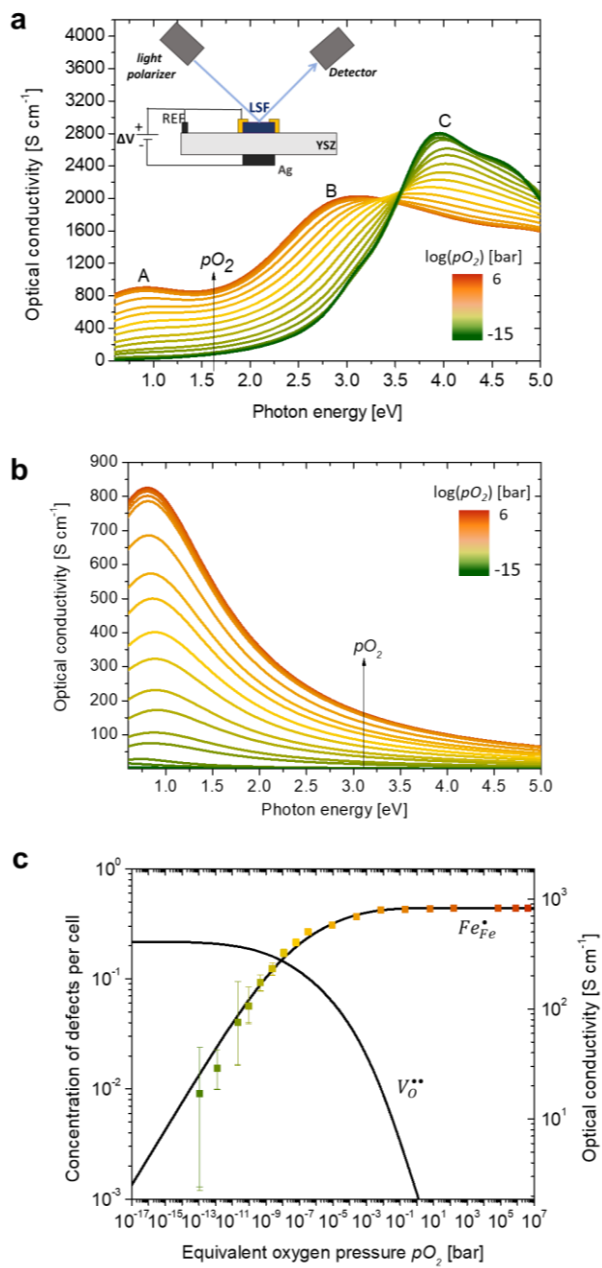

**Figure 3** a) Optical conductivity spectra of the LSF50 thin film measured at 400 °C as a function of equivalent oxygen pressure. The inset shows the scheme of the in-situ ellipsometry measurements: an electrochemical potential is applied across the YSZ electrolyte, varying the chemical potential in the LSF thin films. The variation of the optical properties are detected by ellipsometry. b) Evolution of the low-energy transition A with equivalent oxygen pressure. c) Variation of the maximum of the optical conductivity of the transition A and the concentration of $Fe^{4+}$ per unit cell with the equivalent oxygen pressure (dots), the solid lines represent the fitting of the concentration of $Fe^{4+}$ holes and oxygen vacancies considering a $K_{ox}^{di}$=7500.

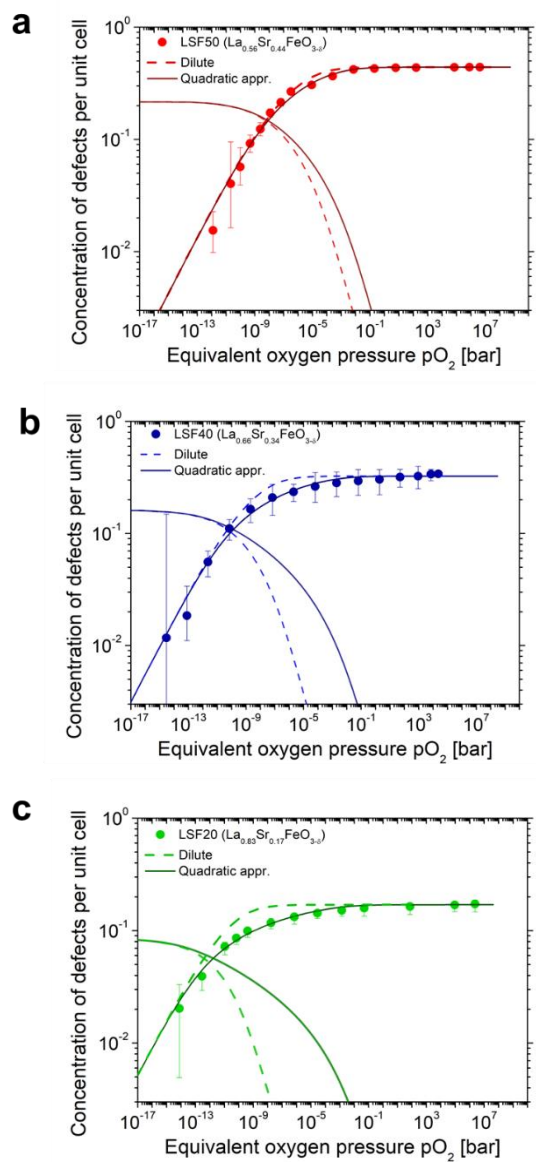

**Figure 4.** Dilute defect chemistry model (dash lines) and quadratic (solid lines) approximation of the concentration of the point defects (symbols) for the a) LSF50, b) LSF40 and c) LSF20 thin films as a function of equivalent oxygen pressure at 400°C.

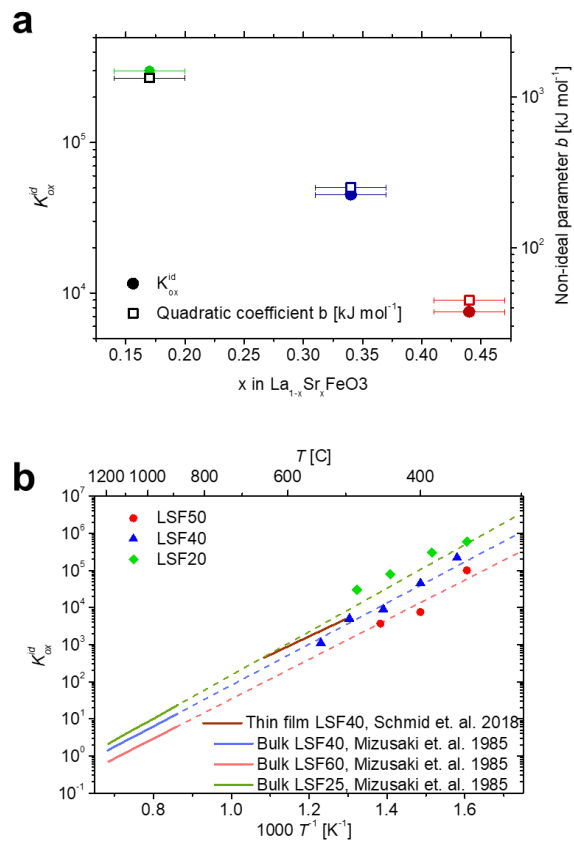

**Figure 5.** a) Equilibrium constant $K_{ox}^{id}$ and quadratic non-ideal parameter b obtained at 400 ºC as a function of the Sr content in the thin films. The dashed lines are intended as a guide for the eye. b) Equilibrium constant $K_{ox}^{id}$ for the oxygen incorporation reaction of the LSF50, LSF40 and LSF20 thin films as a function of temperature (symbols) in comparison with the literature data (solid lines).[46,56] The dashed lines represent the low temperature extrapolation from the literature.

Yunqing Tang, Francesco Chiabrera,[*] Alex Morata, Iñigo Garbayo, Nerea Alayo, Albert Tarancón[*]

**Pushing the Quantification of Point Defects in Thin Film Ferrites to Low Temperatures Using In-situ Ellipsometry**

**TOC:**

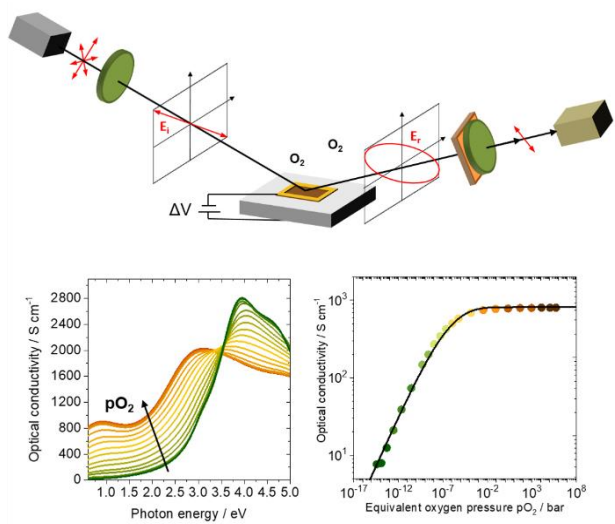

# Supporting Information

**Pushing the Study of Point Defects in Thin Film Ferrites to Low Temperatures Using In-situ Ellipsometry**

*Yunqing Tang, Francesco Chiabrera,* Alex Morata, Iñigo Garbayo, Nerea Alayo, Albert Tarancón**

Y. Tang, Dr. F. Chiabrera, Dr. A. Morata, Dr. Iñigo Garbayo, Dr. N. Alayo, Prof. A. Tarancón
Department of Advanced Materials for Energy Applications, Catalonia Institute for Energy Research (IREC), Jardins de les Dones de Negre 1, 08930, Sant Adrià del Besòs, Barcelona, Spain.
Dr. Iñigo Garbayo, Centre for Cooperative Research on Alternative Energies (CIC energiGUNE), Basque Research and Technology Alliance (BRTA), Alava Technology Park, Albert Einstein 48, 01510 Vitoria-Gasteiz, Spain
Prof. A. Tarancón
ICREA, Passeig Lluís Companys 23, 08010, Barcelona, Spain.

E-mail: f.chiabrera@gmail.com, atarancon@irec.cat

**S.1. Microstructural analysis**

**Figure S1a** shows the XRD diffractogram of the as-deposited LaFeO$_3$ (LFO) thin film on top of a Sapphire (0001) substrate. The LFO thin film shows a pure phase with a pseudo-cubic polycrystalline structure. The out-of-plane lattice parameter obtained from the XRD results as plotted in **Figure S1b** confirms the contraction of the pseudo-cubic LSF structure with the increase Sr content endorsing the behaviour reported in literature. [1–5] The unidentical oxidation state of the as-deposited films leads to the different Fe$^{4+}$ holes concentration that explains the discrepancy of the obtained lattice parameter from a linear approximation. Therefore, the LFO thin film is suitable for studying the holes effects on the optical conductivity of LSF.

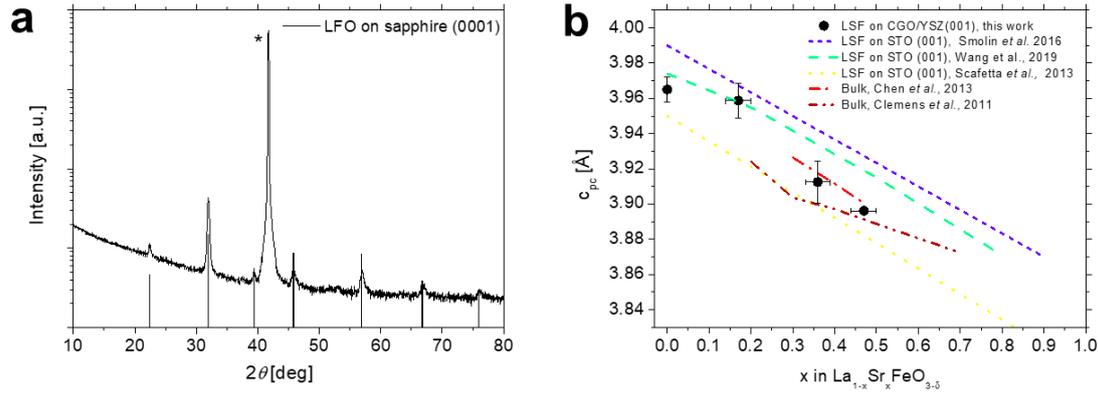

**Figure S1.** a) XRD pattern of the LaFeO$_3$ (LFO) thin film deposited on the (0001)-oriented Sapphire substrate. The LFO phase was identified and the asterisk represents the diffraction peak of the substrate. b) Comparison of the of out-of-plane lattice parameter of LSF thin films as a function of Sr content with literature. [1–5]

**S.2. Ellipsometry data fitting procedure**

The ellipsometry raw data are modeled and fitted using the DeltaPsi2 software, the dielectric properties of LSF thin films are modeled using 5 Lorentzian oscillators as documented in literature:[3]

$$\varepsilon = \varepsilon_0 + \sum_{j=1}^{n} \frac{f_j \omega_{0,j}^2}{\omega_{0,j}^2 - \omega^2 + i\gamma_j \omega} \tag{S1}$$

Where $\varepsilon_0$ is the high frequency dielectric constant, $\omega$ is the angular frequency of the light beam, $\omega_{0,j}$, $f_j$ and $\gamma_j$ are the resonance frequency, intensity and the broadening of the corresponding oscillator.

The inset of **Figure S2a** shows the model used for the fitting consisting of 4 layers: the one side polished substrate, CGO interlayer, LSF thin film and a top layer consisting of 50% vol. of Void and 50% vol. of LSF thin film, considered to be the roughness of the sample (typically < 4 nm). To extract the optical properties of LSF from the raw data, the previously measured and fitted CGO and YSZ optical spectra are used for the fitting. During the fitting procedure the roughness and the thickness of LSF thin film are kept free. The ellipsometry data fitting gives sensitivity to all of parameters included, both those of the oscillators and the thickness of the materials;

thereby, the errors of the optical conductivity presented in this work were calculated by the sensitivity of the parameters of the corresponding oscillator given by the DeltaPsi2 software during the data fitting procedure. Figure S2a shows a high-quality fitting of the ellipsometry data of LSF thin film with an error $\chi^2$=0.04. Once the optical constants of LSF thin film were extracted, the optical conductivity can be calculated:

$$\sigma_{optical} = \frac{4\pi nk}{\lambda Z} \quad \quad (S2)$$

Where $n$ is the refractive index, $k$ is the extinction coefficient, $\lambda$ is the wavelength and $Z$ is a physical constant referring to the impedance of free space.

The extracted optical conductivity of the exemplified LSF50 thin film is shown in **Figure S2b**, as well as the five Lorentzian oscillators used for the modelling.

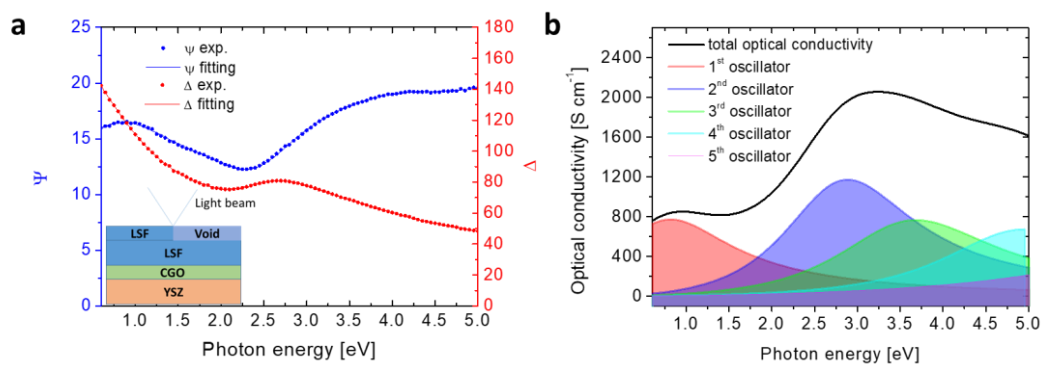

**Figure S2.** a) Raw ellipsometry data (symbols) and fitting (solid lines) for the LSF50 sample recorded at 400 ºC. The inset shows the model used for the data fitting. b) The extracted optical conductivity spectra and the deconvoluted oscillators of the LSF50 thin film.

S.3. Homogeneity of the thin films

In this section, the homogeneity of the thin films will be discussed. Firstly, the cross-section image shown in Figure S3a confirms the high density of the thin films deposited by PLD. Nevertheless, it is well known that around grain boundaries (GBs) or dislocations may present a different composition than the grain bulk, which may lead to changes in the overall optical properties measured by ellipsometry. Generally speaking, grain boundaries in oxides are expected to present an accumulation of oxygen vacancies in the grain core along with a depletion of positive defects concentration in the grain boundaries´ surroundings for maintaining charge neutrality.[6] Due to the high Sr dopant concentration of our LSF thin films, the extension of the space charge will probably be very small (small Debeye length), and mainly composed by hole depletion (most numerous positive defect). Therefore, the effect of GBs could be simulated by an effective medium composed by 95% of bulk LSF and 5% of fully reduced GBs (considering an average grain size of 40 nm and a GB width of 1 nm). We performed a set of simulation of the Δ and Ψ ellipsometry angles expected in our system, where the GB optical properties are set equal to the fully reduced LSF50 layer and varying the optical properties of the bulk from fully oxidized to fully reduced (see **Figure S3**). From the simulations, it is possible to observe that small differences are observed in Ψ angle in the fully oxidized samples but these tends to reduce while reducing the thin films. As commented in the main text, these differences may produce small uncertainties in the exact quantification of the transition A in the fully oxidized thin films at RT, which may vary its linearity depending on the film´s structure. Nevertheless, the effect of GBs is expected to be of minor importance in the in-situ electrochemical analysis of LSF thin films, since a minor variation is observed between the inhomogeneous and heterogeneous model while reducing the sample. For this reason, the quantification of the defect chemistry based on the Brower analysis in LSF thin films still holds its validity.

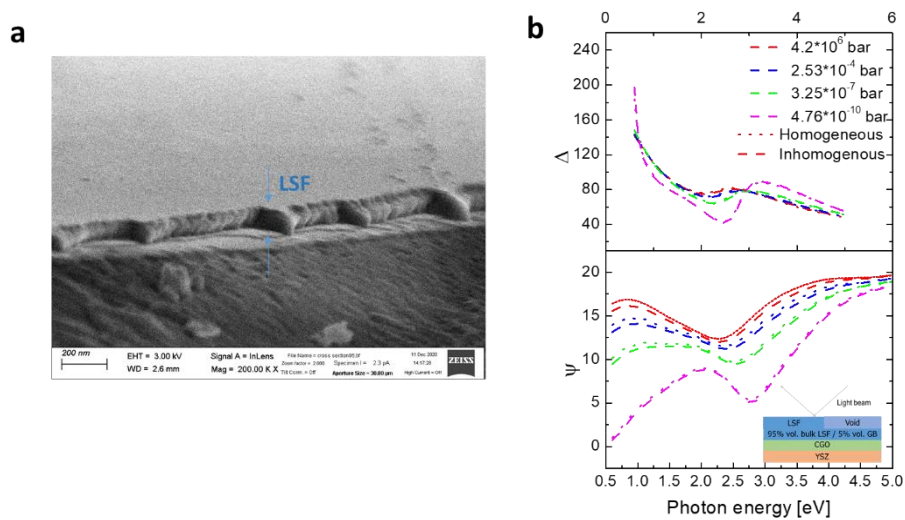

**Figure S3.** a) Cross-section morphology of the LSF50 thin film characterized by SEM. b) Simulation of optical parameters for the homogeneous and inhomogeneous LSF50 sample at 400 °C, the inset shows the model used for simulating the inhomogeneous sample.

**S.4. Temperature effects on the optical conductivity of LSF**

A linear relationship between the maximum of the optical conductivity and the concentration of holes in the LSF system has been discussed in the **section 2.2** of the main text. As commented there, the different slopes observed is originated from the varying temperature. In order to the study the temperature effects on the optical conductivity of LSF, the LSF sample was electrochemically oxidized (see experimental section for further details) at 540°C, then the sample was cooled down to room temperature maintaining the anodic voltage (ensuring a fully oxidized state). The optical properties of the sample were characterized by ellipsometry at each temperature.

**Figure S4a** shows the extracted optical conductivity spectra of the LSF50 thin film recorded at different temperatures as an example. As expected, the spectral weight of the transition A gradually strengthens with sample cooling. Moreover, the transition B can be assigned to a direct transition due to the temperature independent optical conductivity edge. The plotting of the maximum of the optical conductivity of the extracted transition A is shown in **Figure S4b,** confirming the visual observation. A similar behaviour documented in literature associates the temperature effect to the increased phonon absorption and the modified electronic structure.[7] In summary, the temperature is a factor that affects the electronic band structure of the semiconductor-like materials causing the change of the slope of the experimental results.

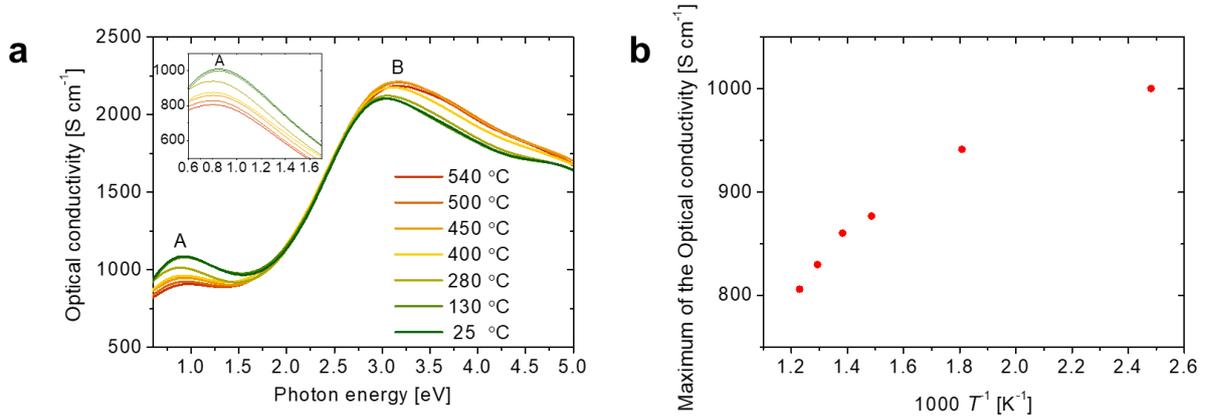

**Figure S4.** a) Optical conductivity spectra of the fully oxidized LSF50 thin film measured at different temperatures, the inset shows the extracted oscillator at the optical transition A. b) Evolution of the maximum of the optical conductivity at the optical transition A as a function of temperature.

**S.5. Electrochemical characterization and equivalent oxygen partial pressure estimation**

The equivalent oxygen partial pressure ($p_{O_2}^{LSF}$) is calculated in the in-situ electrochemical experiments according to the Nernst potential:[8-10]

$$p_{O_2}^{LSF} = p_{O_2}^{RE} \cdot \exp(\frac{4e\Delta E}{kT}) \quad \text{(S3)}$$

Where $p_{O_2}^{RE}$ is the oxygen partial pressure of the environment, $k$ is the Boltzmann constant and $e$ is the electron charge. The overpotential $\Delta E$ at the LSF electrode can be calculated by subtracting the resistive contributions of the YSZ substrates ($R_{YSZ}$) and the silver counter electrode ($R_{Ag}$) to the applied electrochemical bias (**Figure S5a**), as:

$$\Delta E = V_{DC} - R_{YSZ} \cdot I_{DC} - R_{Ag} \cdot I_{DC} \quad \text{(S4)}$$

An anodic Nernstian voltage gives rise to an electrical current from the LSF electrode to the counter electrode (and opposed, in the case of a cathodic voltage), passing through the YSZ substrate. Considering the similarity of the electrochemical behaviour of the materials studied, the LSF50 sample measured at 400 ºC is taken as a representative example; **Figure S5b** shows the measured electrical current as a function of the Nernstian voltage. Impedance spectroscopy measurements were carried out at each potential to measure the resistive contributions in series to the LSF electrode (see **Figure S5c**). All the spectra show a high frequency contribution, associated to the ionic conduction in the YSZ electrolyte, and a middle arc, related to the electrochemical contribution of the Ag paste, and a very resistive low frequency contribution, associated to the incorporation/evolution of the oxygen on the LSF electrode. One can notice that, due to the low range of temperature investigated in this work, the resistive contribution associated to the oxygen incorporation in the LSF layer is very high, shifting the characteristic time of the process below the measurable frequency. This highlights the limitations of the chemical capacitance method for deriving the defect chemistry of oxides at low temperature, where the capacitance associated to the electrode cannot be fully resolved.

**Figure S5d** shows the evolution of the ellipsometry parameters $\Delta$ and $\Psi$ measured for the LSF50 samples at 400 ºC. A monotonic modification of the complex reflectance of the sample is observed varying the bias applied, especially in the low energy region, where interference oscillation typical of semi-transparent layers appears at large cathodic potential. The raw experimental ellipsometry data were fitted with the model described in **Section S.2** to obtain the optical conductivity as a function of the equivalent oxygen partial pressure.

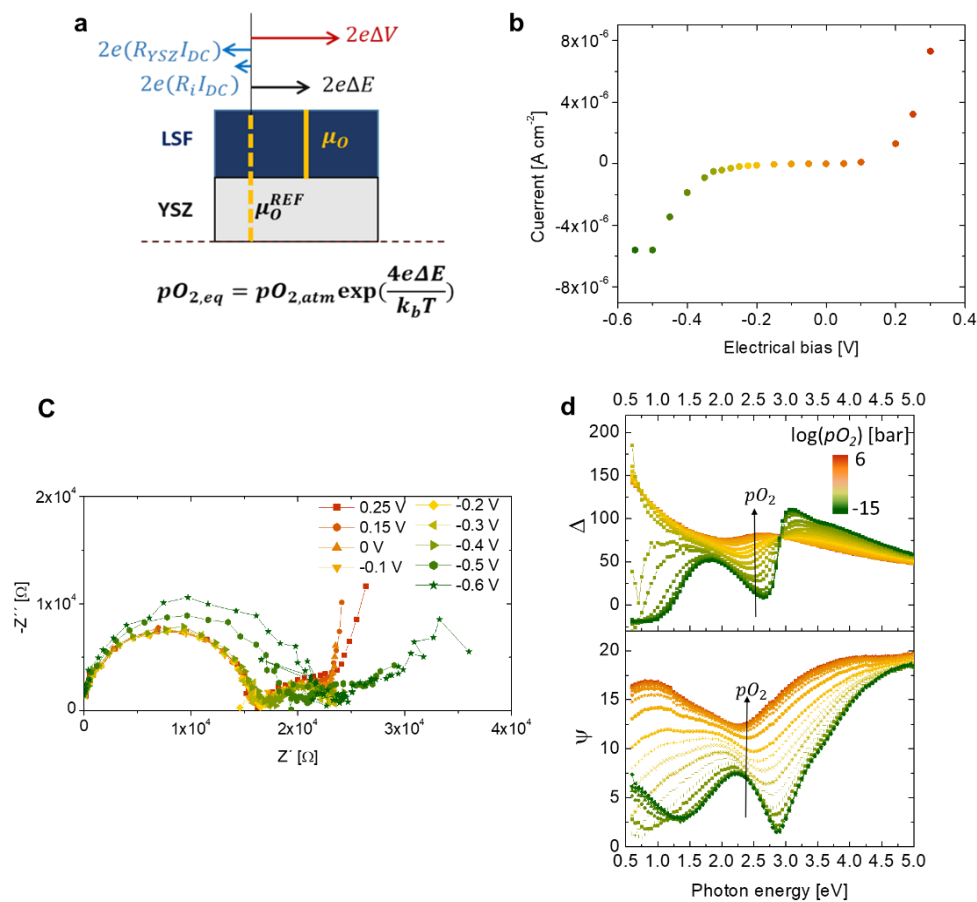

**Figure S5.** a) Scheme of the oxygen chemical capacitance and the resistive contributions in the sample. b) Electrical current of the LSF50 sample induced by the applied Nernstian potential measured at 400 °C. c) Nyquist plot of the representative electrochemical impedance spectra of the LSF50 sample recorded at 400 °C and at different applied electrochemical bias. d) Ellipsometry complex reflectance measured at 400 °C for the LSF50 sample as a function of equivalent oxygen pressure (see the colour scale bar). The arrows indicate the increment of the equivalent oxygen pressure.

In order to study the reversibility of this electrochemical approach, a certain voltage bias was applied after the complete electrochemical oxidation and reduction of the sample, respectively. The optical properties were characterized by ellipsometry (**Figure S6**). The good coincidence of the optical response after the two cycles of measurements gives evidence to the reversibility of the optical response to the redox reactions of the sample.

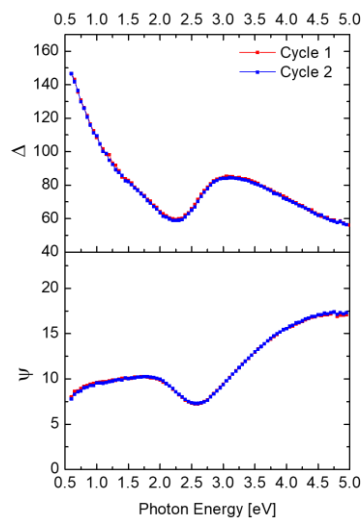

**Figure S6.** Ellipsometry complex reflectance measured at 400 ºC and equivalent oxygen pressure of 0.0087 bar after the full oxidation (cycle 1) and reduction (cycle 2) for the LSF40 sample.

Still, as pointed out in literature,[8-10] three conditions must be fulfilled in order to assure a homogenous oxygen chemical potential in the LSF layers: (i.) Oxygen transport must not be limiting across the LSF/electrolyte interface; (ii.) The oxygen incorporation on the LSF surface is limiting the overall electrode reaction (facile oxygen diffusion); (iii.) The sheet electronic resistance of the LSF layer is low and the electric potential drop along the film's sides are negligible. The use of a CGO barrier avoids the formation of secondary phases at the LSF/YSZ interface, therefore assuring negligible potential drops across this boundary. Condition (ii.) is generally assumed to be true for LSF thin films, where the surface exchange reaction is dominating over the oxygen diffusion across the layer. [9,11,12] The high electronic conductivity of LSF in oxidizing conditions ensure the fulfilment of condition (iii.) in anodic and small cathodic electrochemical potentials.[8-10] However, high cathodic potentials may give rise to a non-homogenous electrochemical potential in the thin films, since the decrease of the hole concentration is expected to progressively decrease the electronic conductivity of LSF layers.

In order to take in account this limitation, a method based on the observation of resistive contribution of YSZ by EIS is considered. Briefly, at high cathodic potential the electrolyte resistance is observed to start to increase, which is related to a loss of electrical equipotential condition in the LSF layer. For this reason, the maximum cathodic potential considered in this work is the one that originates a change in the YSZ resistive contributions, fulfilling condition (iii.). The results were also confirmed by a finite element method (FEM) model for predicting the thin films behaviour under different electronic sheet resistance and oxygen incorporation resistance (see next section, **section S.5.1**).

**S.5.1 FEM simulation of finite sheet resistance in LSF thin films electrodes**

To understand the electronic homogeneity in LSF thin films, the distribution of the electrochemical potential along the sample was modelled by a finite element method (FEM) using COMSOL Multiphysics Modelling software (see **Figure S7**). The model is based on the work of Lynch *et al.*,[13] who simulated the electrochemical distribution in thin film electrodes. One must note that these simulations are mainly intended to depict the effect of inhomogeneous electrochemical potential on the ionic conduction of the YSZ substrates and not to describe the real electrochemical conditions of the thin films under study.

As shown in **Figure S7a** under electrochemical conditions, the reaction of oxygen incorporation competes with the in-plane electronic conduction. Electrons flow from the metallic current collector parallel to the film, reaching the oxygen species on the surface of the electrode were the oxygen reactions take place. It is possible to distinguish two type of currents, the electronic current ($i_e$) along the film length and the ionic current ($i_i$), flowing from the YSZ electrolyte, perpendicular to the LSF film. The geometry used in the model is shown in **Figure S7b** and **Figure S7c**. The electronic current in the LSF thin film is simulated with the Electric Current, shell model, which assume negligible ohmic losses perpendicular to the layer (dependent variable: electronic potential $\Delta\Phi_e$). A fixed potential is applied on the borders of the electrodes, simulating the negligible ohmic resistance of the gold current collector. The ionic current in the YSZ electrolyte is simulated by the electric current module (dependent variable: ionic potential $\Delta\Phi_i$). The two modules are connected by imposing the same normal current density entering/exit the electrode surface $i_{orr} = (\Delta\Phi_e - \Delta\Phi_i)/\tilde{R}_{pol}$, where $\tilde{R}_{pol}$ is the polarization resistance of the LSF electrode. In the model, LSF is considered surface limited (infinite oxygen diffusion in the thin film) and the ionic conductivity of YSZ is fixed at $10^{-4}$ S/cm.

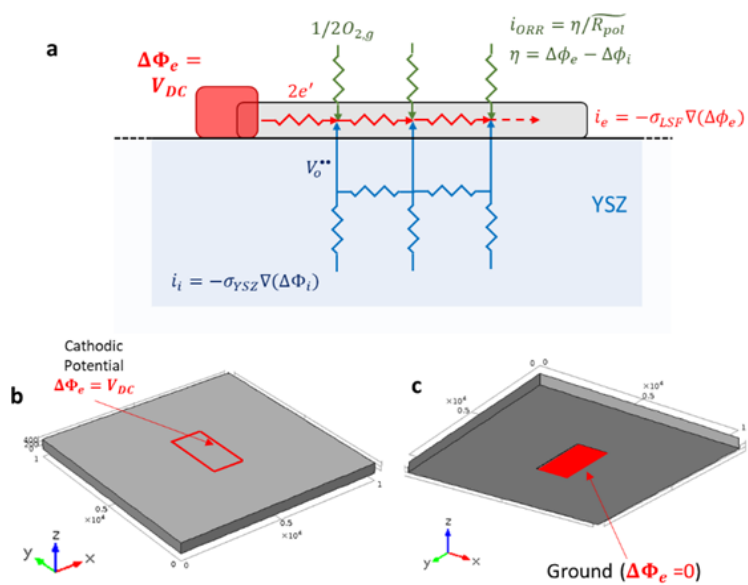

**Figure S7.** a) Schematic representation of the Nernstian voltage driven electronic conduction and oxygen incorporation reaction in the samples. Schemes of voltage distribution at b) LSF lectrode and c) silver counter electrode under an anodic potential.

**Figure S8** shows the ionic potential in the YSZ electrolyte simulated varying the sheet resistivity of the electrode ($10^{-1}$ to $10^2$ ohm*cm) and the polarization resistance of the surface ($10^2$-$10^5$ ohm*cm$^2$). These values were chosen to cover realistic values under the experimental conditions studied. The results show that low values of sheet resistivity and high values of surface polarization resistances ensures equipotential electrodes thin film. Nevertheless, it is also interesting to note that equipotential layers can be obtained even with relatively large values of electronic resistivity, provided that the surface polarization resistance is high enough. This condition is easily observed at low temperature, due to the higher activation energy of oxygen incorporation (1 to 2 eV) compared to electronic transport (0.2 eV).

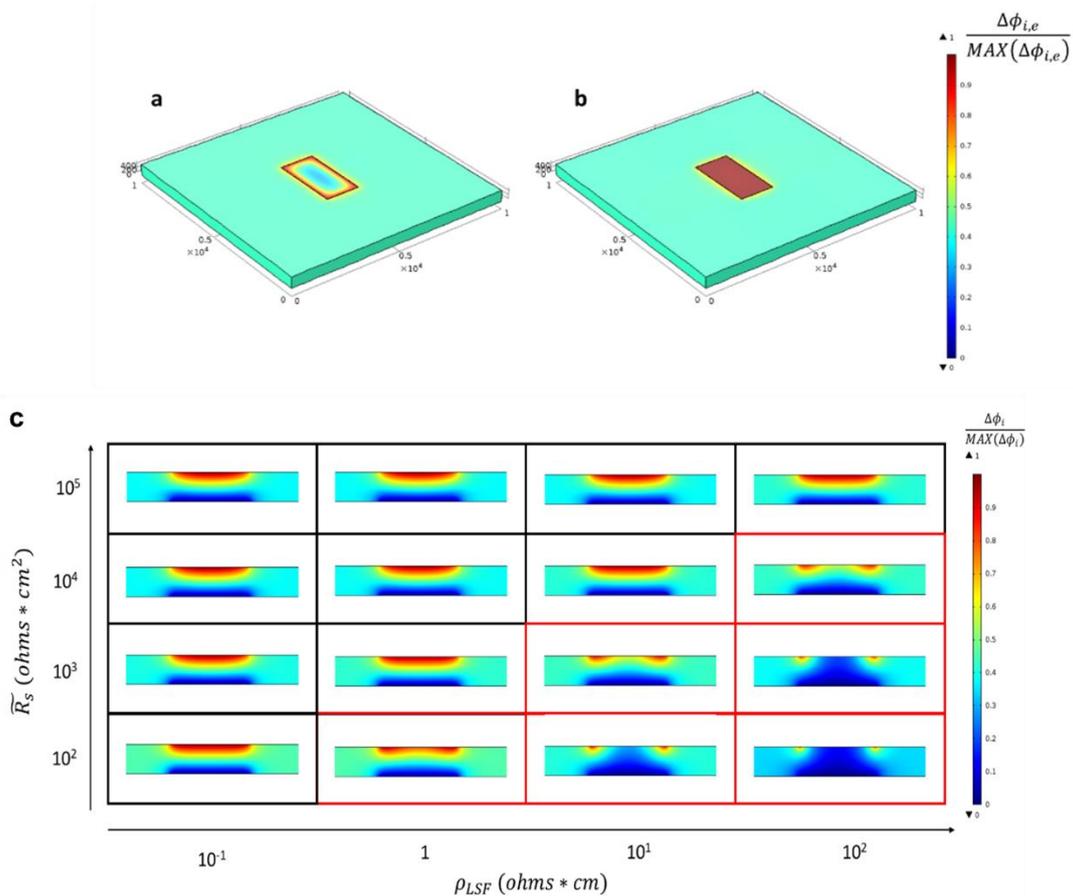

**Figure S8.** a) Heterogeneity and b) homogeneity of electronic distribution in LSF thin film. c) Modelling of the electronic distribution in the sample as a function of the sheet resistivity of LSF film ($\rho_{LSF}$) and the oxygen incorporation resistivity of the YSZ electrolyte ($\tilde{R}_s$).

Nevertheless, as commented before, describing the real behavior of the LSF electrodes is challenging, since both surface polarization and electronic conductivity of the layer vary under electrochemical bias. For this reason, we considered a different approach based on the observation of the YSZ resistance in impedance spectroscopy under different chemical potential. **Figure S9a** shows a representative example of the LSF50 sample measured at 400 ºC. The high frequency electrochemical element, associated with the ionic conduction in the YSZ electrolyte, is constant for any value of bias down to -0.4 V. Below this value, the resistance associated with this element starts to increase. The origin of this variation is related to the decrease of active area when passing from homogenous to inhomogeneous chemical potential conditions, as depicted in **Figure S9b** and **Figure S9c**. In other words, for low cathodic potential (V > -0.4 V), the low electronic resistivity of LSF thin films (and the high polarization resistance) ensure an equipotential chemical potential along the LSF thin film. Increasing the cathodic potential (V < -0.4 V), the reduction of the hole concentration give rise to an increase of electronic resistivity that, linked also to a reduction of polarization resistance, give rise to a non-homogenous distribution of the ionic potential in the LSF layer. This change of regime is observed in the ionic contribution of the YSZ in impedance spectroscopy, where an increase of resistance is measured.

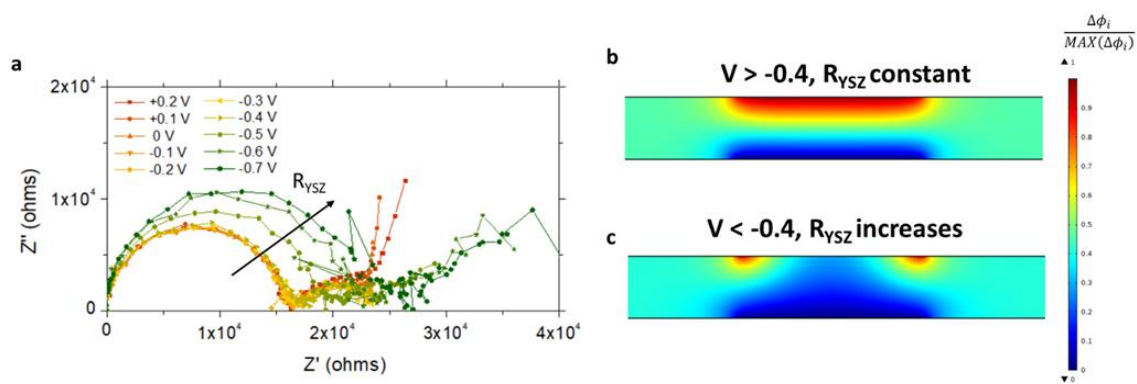

**Figure S9.** Representative electrochemical impedance spectra of the LSF50 sample measured at 400 ºC as a function of Nernstian voltage bias. The raw indicates the increase of the ionic resistance of YSZ. b) At low cathodic voltage where the ionic resistance is still constant, the LSF thin film is equipotential. c) At high cathodic voltage where the ionic resistance of YSZ starts to increase, the LSF thin film is not equipotential.

**S.6. Chemical expansion or contraction of LSF**

The ellipsometry technique is not only sensitive to the optical constants of LSF films but also sensitive to the variation of the thickness of the thin film as a function of equivalent $pO_2$. **Figure S10** shows the variation of the thickness of the LSF50 thin film with the equivalent $pO_2$ given by the ellipsometry data fitting. In order to compare the results with the literature, the second order polynomial expression developed by Chen *et al.* was considered,[4] where the out of plane lattice expansion was measured as a function of the oxygen vacancies concentration in the system. In **Figure S10,** the lattice expansion predicted by this model is reported. Please note that we consider the oxygen vacancies concentration calculated by the fitting of optical conductivity (**Figure 3c** in the main text), meaning that the thickness evolution measured and the model results are independent. Considering the margin of error, the trend of the experimental results are in rough agreement with literature.

It must be note that the large variation of optical constants at high anodic and cathodic voltage decreases the sensitivity of the thickness measurements lowering the confidence of the fitting in the same way for the other materials LSF40 and LSF20 studied in this work. For this reason, in this work the variation of optical conductivity is preferred to quantify the point defect concentration in the LSF system. Nevertheless, the sensitivity of ellipsometry to the thickness measurements may allow tracking the defect chemistry in oxide thin films were no relevant changes of optical properties take place with $pO_2$.

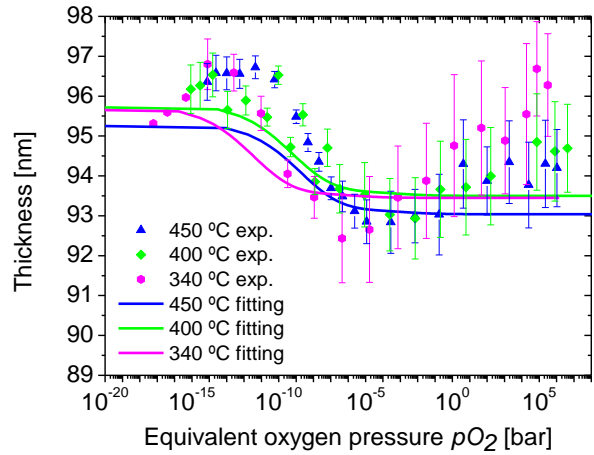

**Figure S10.** Thickness of the LSF50 thin film as a function of equivalent *pO₂* at different temperatures obtained from the ellipsometry data fitting (symbols). Solid lines shows the thickness variation expected using a second-order polynomial model developed in literature[4] and the concentration of oxygen vacancies calculated by optical conductivity (**Fig. 3c** in the main text).

**S.7. Defect chemistry modelling**

Non-dilute behaviour can be described by considering a non-diluted chemical potential of defect species:

$$\mu_i = \mu_i^0 + RTln([i] \cdot \gamma_i) \tag{S5}$$

Where $\mu_i^0$ is the standard chemical potential, $[i]$ is the concentration of the *i* specie and $\gamma_i$ is the activity coefficients, that may deviate from the regular unity. The oxygen incorporation equilibrium constant (Equation (3) in the main text) then becomes:

$$K_{ox} = \frac{[Fe_{Fe}^\bullet]^2 \cdot [O_O^\times]}{(pO_2)^{\frac{1}{2}} \cdot [V_O^{\bullet\bullet}] \cdot [Fe_{Fe}^\times]^2} \cdot \frac{\gamma_{Fe_{Fe}^\bullet}^2 \gamma_{O_O^\times}}{\gamma_{Fe_{Fe}^\times}^2 \gamma_{V_O^{\bullet\bullet}}} = K_{ox}^{id} \cdot \frac{\gamma_{Fe_{Fe}^\bullet}^2 \gamma_{O_O^\times}}{\gamma_{Fe_{Fe}^\times}^2 \gamma_{V_O^{\bullet\bullet}}} \tag{S6}$$

One can note that the equilibrium constant is now composed by the product of the dilute $K_{ox}^{id}$ and the activity coefficients, not necessarily constant in the whole *pO₂* range. In order to better define the problem, it is useful to consider the definition of Gibbs free energy of the oxygen incorporation reaction, as:

$$\Delta G_{ox} = -RTln(K_{ox}) = \Delta G_{ox}^{id} + \Delta G_{ox}^{ex} \tag{S7}$$

In this way, the energetics of oxygen incorporation reaction (Equation (2) in the main text) are defined by the standard term ($\Delta G_{ox}^{id} = -RTln(K_{ox}^{id})$, constant for any $pO_2$) and the activity term ($\Delta G_{ox}^{ex} = -RTln(\frac{\gamma_{Fe_{Fe}^{\bullet}}^2 \gamma_{O_O^\times}}{\gamma_{Fe_{Fe}^\times}^2 \gamma_{v_O^{\bullet\bullet}}})$), representing the deviation from the standard free energy of the ideal solution, that is the driving energy for the modification of the oxygen incorporation equilibrium. As commented in the main text, a general solution of this problem was proposed by Mizusaki et al., who considered $\Delta G_{ox}^{ex}$ to be linearly proportional to the point defect concentration, as:[14]

$$\Delta G_{ox}^{ex} = a[Fe^\bullet] \tag{S8}$$

Alternatively, in this work also a quadratic approximation is considered, probably originated by a non-negligible interaction between the point defect concentration:

$$\Delta G_{ox}^{ex} = b[Fe^\bullet]^2 \tag{S9}$$

**Figure S11** shows the comparison of dilute, linear and quadratic models. All of the three models fit well the holes concentration at high $pO_2$ and the three models fall into the same value at low pressure. The major difference between these three models is observed at the intermediate $pO_2$, where the experimental holes concentration is lower than that expected by the dilute model, and a deviation of the experimental results from the fitting curve is clearly observed. The non-dilute model in which the Gibbs free energy of the oxygen incorporation reaction is considered to be linearly proportional to the holes concentration[14] enhances the fitting although the experimental results are still more reduced than that expected by this model. In this sense, we employed a quadratic model in which the Gibbs free energy of the oxygen incorporation reaction is a second-order polynomial function of the holes concentration. One can note that this quadratic model fits well the holes concentration. Therefore, the quadratic model is used for fitting the holes concentration as a function of $pO_2$ in this study.

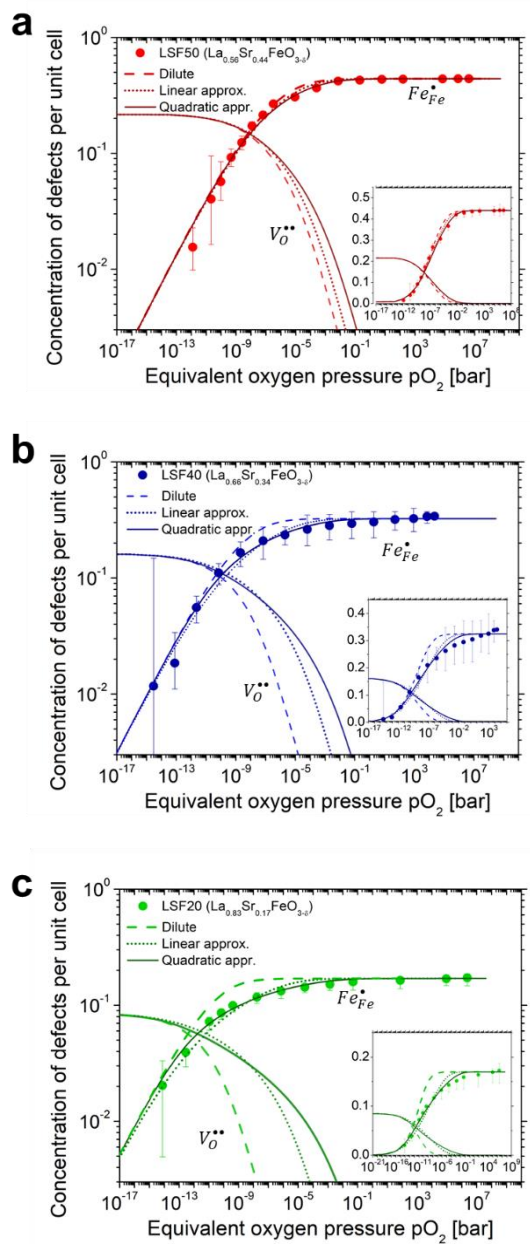

**Figure S11.** Dilute defect chemistry model (dash lines), linear approximation (dot lines) and quadratic approximation (solid lines) of holes concentration $Fe_{Fe}^{\cdot}$ and concentration of oxygen vacancies $V_{\ddot{O}}$ for fitting the experimental holes concentration (symbols) as a function of equivalent $pO_2$ for a) LSF50 b) LSF40 and c) LSF20 thin films.

### S.8. In situ ellipsometry measurements as a function of temperature

The in situ ellipsometry measurements were carried out at different temperatures. **Figure S12** shows a similar variation of holes concentration as a function of equivalent $pO_2$ at different temperatures, a shift of the curves of holes concentration is clearly observed due to the temperature effects. The equilibrium constant $k_{ox}^{id}$ as a function of temperature is obtained from the fitting for the holes concentration using the quadratic approximation.

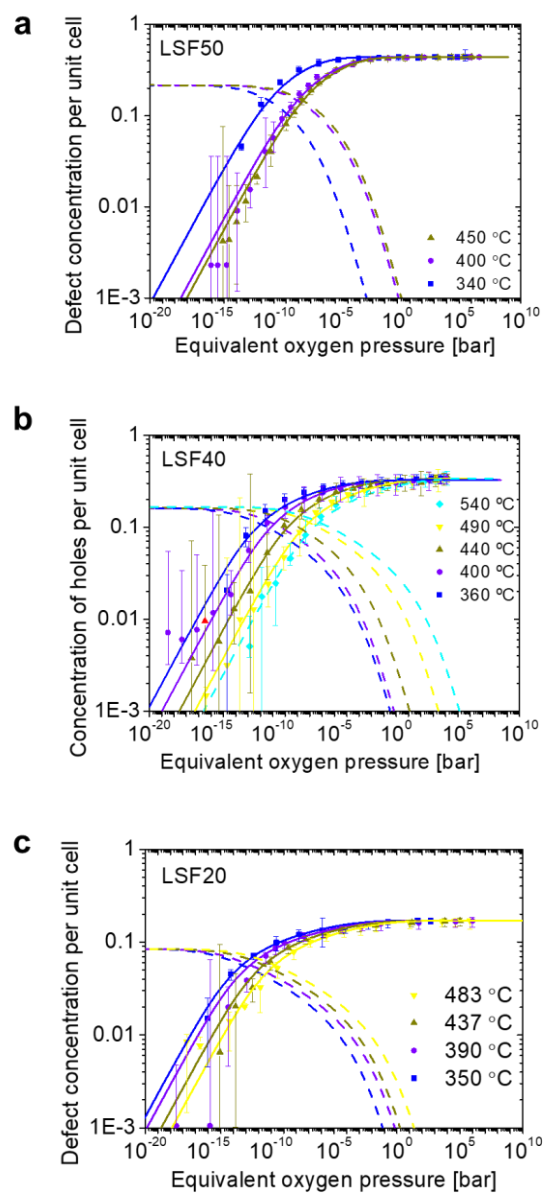

**Figure S12.** Experimental concentration of holes (symbols) as a function of equivalent $pO_2$ with the quadratic approximation of $Fe^{4+}$ holes (solid lines) and oxygen vacancies (dash lines) for a) LSF50, b) LSF40 and c) LSF20 thin films at different temperatures.